\documentclass[aps,pra,showpacs,english,10pt,a4,nofootinbib,notitlepage,twocolumn,superscriptaddress]{revtex4-2}

\usepackage[utf8]{inputenc}
\usepackage[T1]{fontenc}
\usepackage[english]{babel}  
\usepackage{datetime}
\usepackage{dsfont}
\usepackage[colorlinks={},linkcolor=blue,citecolor=blue,urlcolor=blue,pdfauthor={ },pdftitle={ },pdfsubject={ },pdfkeywords={ }]{hyperref}
\usepackage{amssymb,amsmath,amsfonts,amsthm}
\usepackage{mathtools}
\usepackage{verbatim}
\usepackage{enumerate}
\usepackage{graphicx}
\graphicspath{{figs/}}
\usepackage{bbm}
\usepackage{booktabs}

\usepackage[caption=false]{subfig}

\usepackage{color}

\usepackage{xcolor}

\definecolor{dred}{rgb}{.8,0.2,.2}
\definecolor{ddred}{rgb}{.8,0.5,.5}
\definecolor{dblue}{rgb}{.2,0.2,.8}
\definecolor{dgreen}{rgb}{.2,0.5,.2}

\usepackage{graphicx}
\usepackage{float}

\begin{document}
	
\title{Observation of Nonlinear Spin Dynamics in Dual-Cell Atomic Gases}	
\author{Xiaofan Wang}
\affiliation{Zhejiang Provincial Key Laboratory for Quantum Precision Measurement, School of Physics, Zhejiang University of Technology, Hangzhou, 310023, China}

\author{Haitao Lu}
\affiliation{Zhejiang Provincial Key Laboratory for Quantum Precision Measurement, School of Physics, Zhejiang University of Technology, Hangzhou, 310023, China}

\author{Hengyan Wang}
\email{hywang@zust.edu.cn}
\affiliation{Department of Physics, Zhejiang University
of Science and Technology, Hangzhou 310023, China}

\author{Zhihuang Luo}
\email{luozhih5@mail.sysu.edu.cn}
\affiliation{Guangdong Provincial Key Laboratory of Quantum Metrology and Sensing, and School of Physics and Astronomy, Sun Yat-Sen University (Zhuhai Campus), Zhuhai 519082, China.}

\author{Wenqiang Zheng}
\email{wqzheng@zjut.edu.cn}
\affiliation{Zhejiang Provincial Key Laboratory for Quantum Precision Measurement, School of Physics, Zhejiang University of Technology, Hangzhou, 310023, China}


\begin{abstract}
Nonlinear spin systems exhibit rich and exotic dynamical phenomena, offering promising applications ranging from spin masers and time crystals to precision measurement. Recent theoretical work [T. Wang et al., Commun. Phys. 8, 41 (2025)] predicted intriguing nonlinear dynamical phases arising from inhomogeneous magnetic fields and feedback interactions. However, experimental exploration of these predictions remains lacking. Here, we report the observation of nonlinear spin dynamics in dual-bias magnetic fields with dual-cell alkali-metal atomic gases and present three representative stable dynamical behaviors of limit cycles, quasi-periodic orbits, and chaos. Additionally, we probe the nonlinear phase transitions between these phases by varying the feedback gain and the difference of dual-bias magnetic fields. Furthermore, we demonstrate the robustness of the limit cycle and quasi-periodic orbit against the noise of magnetic fields. Our findings establish a versatile platform for exploring complex spin dynamics and open new avenues for the realization of multimode spin masers, time crystals and quasi-crystals, and high-precision magnetometers.
\end{abstract}

\maketitle

\section{Introduction}
Nonlinear effects have been found in various systems and play a significant role in science and engineering~\cite{Strogatz2018Nonlinear}. Specifically in spin systems, nonlinear dynamics is crucial for the prediction of complex spin behaviors such as self-oscillations and chaos~\cite{Jenkins2013Self, Bloom1962Principles, Chalupczak2015Alkali, Bevington2021Object, Robinson1964He3, Yoshimi2002Nuclear, Jiang2021Floquet, Su2022Review, Chupp1994Spin, Stoner1996Demonstration, Bear1998Improved, Sato2018Development, Bevington2020Dual, Feng2025Nonlinear, Chacko2025Multimode, Lin2000Resurrection, Abergel2002Chaotic, Liu2019Trispin}.
The nonlinearity induced by feedback mechanisms causes self-oscillations of the collective spins that persist far beyond the transverse relaxation time $T_2$~\cite{Chupp1994Spin, Yoshimi2002Nuclear}. Such nonlinear oscillations have robust self-organizing patterns in time and ultrahigh-resolution spectra in frequency, promoting the significant development of spin masers \cite{Bloom1962Principles, Chalupczak2015Alkali, Bevington2021Object, Robinson1964He3, Yoshimi2002Nuclear, Jiang2021Floquet, Su2022Review, Chupp1994Spin, Stoner1996Demonstration, Bear1998Improved, Sato2018Development, Bevington2020Dual, Gross1979Maser, Moi1983Rydberg, Goy1983Rydberg} and time crystals~\cite{Choi2017Observation, Liu2024Higher, Wu2024Dissipative, Kongkhambut2022Observation, Greilich2024Robust, Huang2024Observation, Wang2025Observation, Greilich2025Exploring, Liu2025Bifurcation}.
The utilization of dual-species spin masers can further eliminate changes in precession frequency caused by long-term drifts of the magnetic field~\cite{Chupp1994Spin, Stoner1996Demonstration, Bear1998Improved, Sato2018Development, Bevington2020Dual}, which are extremely advantageous for the precision measurement of frequency shifts and enable possible applications in the search for permanent electric dipole moments (EDM) \cite{Rosenberry2001Atomic, Inoue2016Frequency}, and tests of fundamental physics beyond the standard model \cite{, Bear2000Limit, Safronova2018Search, Terrano2022Comgnetometer}.   

Although nonlinear spin dynamics have been observed in various systems such as alkali-metal atoms~\cite{Bloom1962Principles, Chalupczak2015Alkali, Bevington2021Object, Gross1979Maser, Moi1983Rydberg, Goy1983Rydberg}, noble gases~\cite{Robinson1964He3, Yoshimi2002Nuclear, Jiang2021Floquet, Su2022Review}, and nuclear magnetic resonance~\cite{Lin2000Resurrection, Abergel2002Chaotic, Chacko2025Multimode}, most experiments involving single-species spins were conducted in a bias magnetic field and only demonstrated the behaviors of the stable limit cycle which is a closed trajectory in phase space having the property that at least one other trajectory spirals into it as time approaches infinite~\cite{Strogatz2018Nonlinear, Jenkins2013Self}. Even when multiple species of spins with different intrinsic Larmor frequencies were presented in a homogeneous magnetic field, previous studies ~\cite{Chupp1994Spin, Stoner1996Demonstration, Bear1998Improved, Sato2018Development, Bevington2020Dual} typically treated each spin species separately and neglected the critical impact of their interactions on the collective dynamics. 
Recent theoretical work by Wang et al.~\cite{Wang2025Feedback, Wang2025Bifurcations, Wang2025Nonlinear} introduced nonlinear dynamics into the feedback-driven spin systems and predicted that collective spins under an inhomogeneous magnetic field exhibit a richer stable dynamical phase diagram including quasi-periodic orbits and chaos, in addition to the limit cycles observed previously.
These nonlinear dynamical phases hold prospective applications in multimode spin masers, time crystals and quasi-crystals, and high-precision magnetometers. However, experimental observation of these new dynamical phases and their nonlinear phase transitions remains unexplored and challenging.

Here we report the observation of nonlinear spin dynamics in a dual-cell self-oscillating rubidium magnetometer, where two vapor cells experience distinct bias magnetic fields. This configuration naturally introduces two intrinsic Larmor frequencies coupled through a common feedback loop. By tuning the bias-field difference and feedback strength, we experimentally map out the phase diagram and identify three distinct dynamical regimes: synchronized limit cycles, quasi-periodic oscillations, and chaotic trajectories.
In the phase of limit cycles, the atomic spins in two cells exhibit a collective behavior and self-organize to oscillate at a single synchronization frequency. In the phase of quasi-periodic orbits, the spins manage to synchronize and sustain a self-oscillation with multiple incommensurate frequencies. These self-oscillations observed in the experiment, also known as maser oscillations, can persist significantly longer than the transverse relaxation time $T_2$, resulting in very narrow peaks in the spectrum. Consequently, it is an advantage for long-term measurement of frequency shifts. Moreover, in the presence of the magnetic field noise, both the limit cycle and quasi-periodic orbit are robust, which can be considered as the realization of time crystals and quasi-crystals if they spontaneously break the continuous time-translation symmetry.
In the phase of chaos, the dynamical behavior is highly sensitive to initial conditions and can be found to have a close resemblance to the butterfly pattern of the well-known Lorenz equations~\cite{HAKEN197577}.

{\section{Nonlinear Spin Dynamics in a Dual Spin-Ensemble System}
We investigate the nonlinear spin dynamics of two atomic ensembles subjected to a common feedback-driven magnetic field while placed in distinct static magnetic fields. The experimental configuration, illustrated in Fig. \ref{ExpSetup}, consists of two independent thermal $^{87}$Rb vapor cells located in separate static magnetic fields $B^0_{z,1}$ and $B^0_{z,2}$ generated by individual coil systems, enabling independent field control. A pump laser beam is divided into two branches to polarize the atomic spins of each ensemble along the $z$-axis. A single linearly polarized probe beam sequentially traverses both vapor cells, interacting with the ensembles through the Faraday rotation effect. The balanced photodetector converts the modulated signal into an electrical one. The Faraday rotation angle is proportional to the atomic spin component along the light-propagation direction, making the photodetector output proportional to the sum of the $x$-axis spin polarizations of the two ensembles, $
S\left( t \right) \propto M_x\left( t \right) =M_{x,1}\left( t \right) +M_{x,2}\left( t \right) $. The resulting electrical signal is fed back through a variable resistor to a pair of $y$-axis coils surrounding each cell, producing an identical feedback magnetic field for both ensembles. The feedback field is expressed as $B_y(t) = -\alpha {M_x}(t) /\gamma$, where $\gamma$ is the gyromagnetic ratio of $^{87}$Rb and $\alpha$ is the feedback coefficient determining the coupling strength between spin dynamics and the feedback field. The coefficient $\alpha$ can be tuned via the variable resistor. 

Consider the feedback and dual-bias magnetic fields, i.e., \(\mathbf{B_i} = (0, -\alpha {M}_x(t)/\gamma, B^0_{z, i})\) for $i=1, 2$, the system dynamics are described by nonlinear Bloch equations:
\begin{equation}
\begin{cases}
\dfrac{dM_{x, i}}{dt} = \omega_i M_{y, i} + \alpha {M}_x M_{z, i} - \dfrac{M_{x, i}}{T_2} \\[10pt]
\dfrac{dM_{y, i}}{dt} = -\omega_i M_{x, i} - \dfrac{M_{y, i}}{T_2} \\[10pt]
\dfrac{dM_{z, i}}{dt} = -\alpha {M}_x M_{x, i} + \dfrac{M_0 - M_{z, i}}{T_1},
\end{cases}
\label{Bloch}
\end{equation}
where \(\omega_i = \gamma B^0_{z, i}\) represents the Larmor frequency of the $^{87}$Rb spins in the $i$th cell,
\(M_0\) is the equilibrium magnetization, \(T_1\) and \(T_2\) are the longitudinal and transverse relaxation times, respectively. 
Let $\Delta\omega = \omega_1 - \omega_2$ denote the frequency difference between two vapor cells. When the magnetic field is homogeneous ($\Delta \omega = 0$), the system only exhibits the limit-cycle behavior characteristic of a single spin ensemble if $\alpha$ is larger than the critical value $\alpha_c=1/(T_2 M_0)$ \cite{Yoshimi2002Nuclear}.
When $\Delta \omega \neq 0$, the coupled dual-cell spin system exhibits rich nonlinear dynamics, including stable limit cycle, quasi-periodic orbit, and chaos~\cite{Wang2025Feedback}. A limit cycle corresponds to a stable closed trajectory in phase space, where nonlinear coupling synchronizes the two initially distinct Larmor precessions to a single frequency, establishing periodic motion. In contrast, quasi-periodic motion consists of deterministic yet non-repeating trajectories arising from the superposition of incommensurate frequencies. At higher feedback strength, chaotic dynamics emerge, characterized by exponential sensitivity to initial conditions and trajectories confined to a fractal strange attractor, yielding an aperiodic evolution distinct from both periodic and quasi-periodic regimes.
To analyze these behaviors and visualize the evolution trajectories of the dual-cell system, the equations in Eq.~(\ref{Bloch}) are solved numerically. The intrinsic parameters used in the simulation are \(\gamma = 7\) Hz/nT, \(T_1 = 5 \, \text{ms}\), \(T_2 = 2 \, \text{ms}\), \(M_0 = 0.5\).


\begin{figure}
    \centering
    \includegraphics[width=0.5\textwidth]{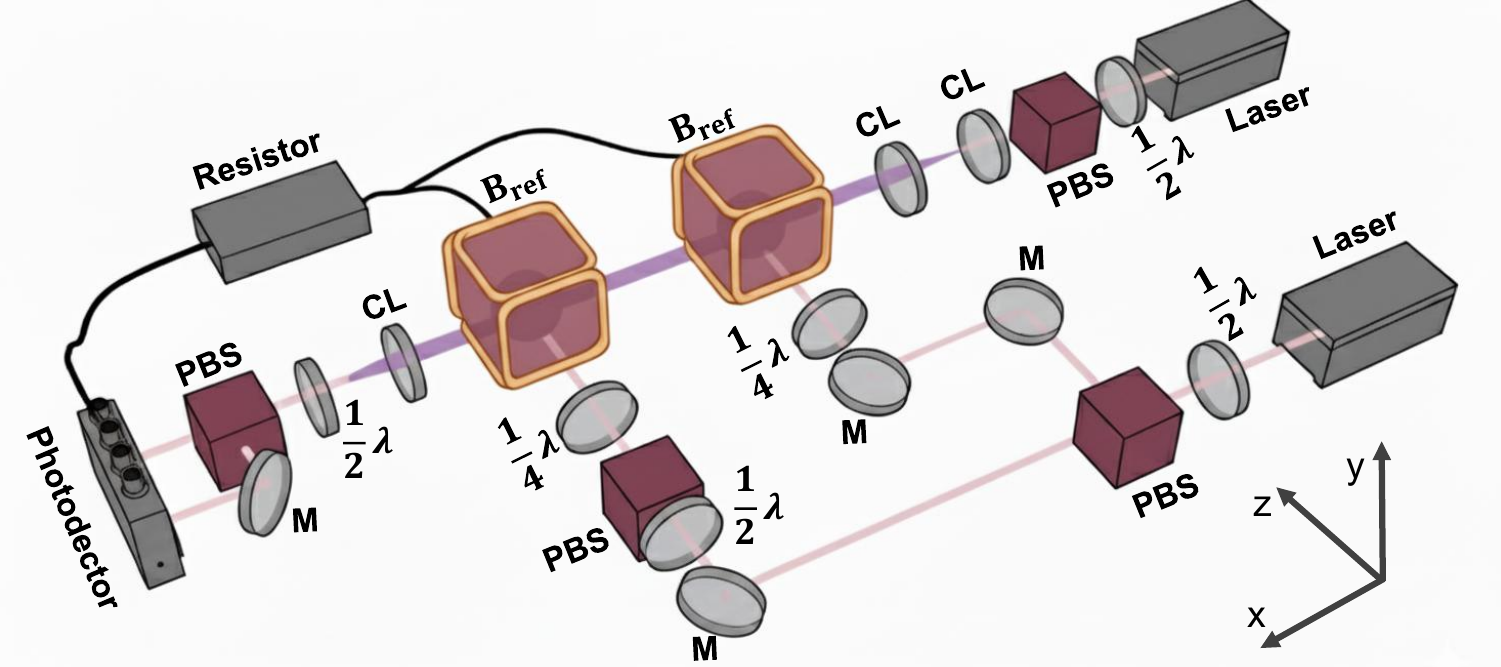} 
    \caption{Experimental setup of the dual-cell self-oscillating magnetometer. The detected probe signal is fed back through coils as $B_{\text{ref}}$ to sustain spin precession and generate nonlinear dynamics. Optical components are labeled as follows: PBS (polarizing beam splitter); M (mirror); CL (convex lens); $\frac{1}{2}\lambda$ and $\frac{1}{4}\lambda$ (half- and quarter-wave plates). The coordinate axes $(x,y,z)$ indicate the laboratory reference frame.}
    \label{ExpSetup}
\end{figure}


We numerically track the time evolution of the total spin polarization components $\{{M}_x, {M}_y, {M}_z\}$. The collection of solutions is visualized in three-dimensional phase space, as shown in Fig.~\ref{NumericalSimFig}. To reveal the dissipative characteristics more intuitively, we introduce a Poincaré section by selecting the plane ${M_y}=0$ and recording the intersection points of the trajectory with this plane. The distribution of these intersection points (red markers in Fig.~\ref{NumericalSimFig}) reflects the dispersion of different dynamical states in phase space.

\begin{figure*}
\centering   \includegraphics[width=0.9\textwidth]{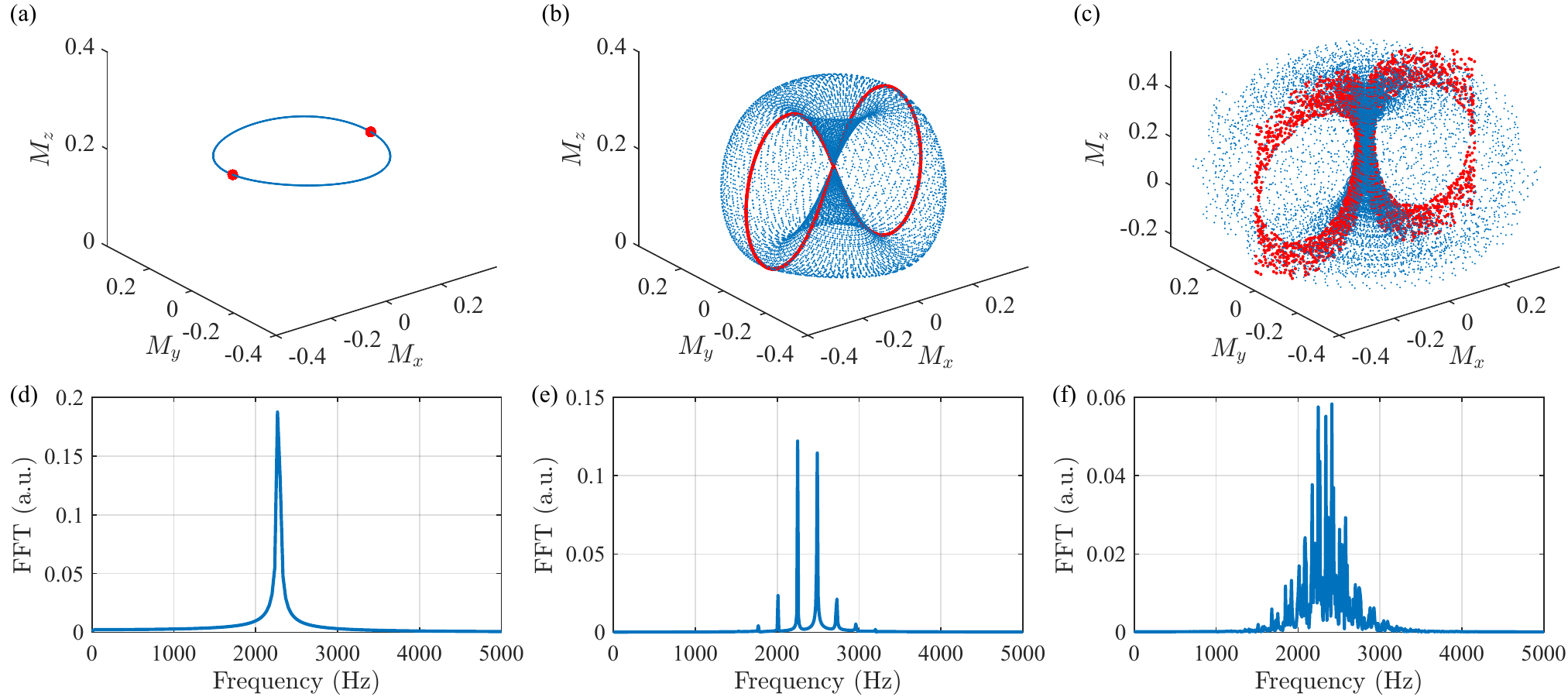}    \caption{Representative trajectories of the average spin polarization $\{\overline{M}_x, \overline{M}_y, \overline{M}_z\}$ in three dynamical phase regimes: (a) limit cycle, (b) quasi-periodic orbit, and (c) chaos. Blue dots denote the simulated trajectories, while red markers highlight the corresponding Poincaré sections.}
\label{NumericalSimFig}
\end{figure*}
When $\Delta\omega$ is small, nonlinear coupling locks the two spin ensembles into a common frequency, resulting in a stable limit-cycle behavior. As shown in Fig.~\ref{NumericalSimFig}(a), for $\Delta\omega = 2\pi\times40$ rad/s, and $\alpha = 16 \alpha_c$, the phase-space trajectory forms a closed loop, while the Poincaré section reduces to two discrete fixed points—signatures of a periodic oscillation. Correspondingly, the Fourier spectrum of the time-domain signal exhibits a single-frequency peak. With increasing $\Delta\omega$ at constant $\alpha$, the coupling weakens and each ensemble oscillates near its own Larmor frequency. The resulting motion arises from the superposition of two incommensurate frequencies (i.e., with an irrational ratio), leading to quasi-periodic evolution. In phase space, the trajectory forms a never-closing loop, whereas the Poincaré section becomes a continuous closed curve, as shown in Fig.~\ref{NumericalSimFig}(b) for $\Delta\omega=2\pi\times220$ rad/s and $\alpha = 16 \alpha_c$. When $\Delta\omega$ is insufficient to completely separate the resonance peaks of the two ensembles, a sufficiently strong feedback coefficient can drive the system into a chaotic regime. Under this condition, the spin dynamics becomes extremely sensitive to initial conditions. For $\alpha = 20\alpha_c$ rad/s and $\Delta\omega = 2\pi\times 110$ rad/s, the numerical solution reveals chaotic motion confined to a strange attractor, as illustrated in Fig.~\ref{NumericalSimFig}(c). The dense, irregular surfaces in phase space indicate strong trajectory divergence under small perturbations, while the structured butterfly-like distribution of points in the Poincaré section resembles the well-known Lorenz attractor—further confirming the presence of chaos.


\section{Experimental observation of Nonlinear Spin Dynamics}
\subsection{Experimental setup}

We then experimentally observe the nonlinear spin dynamics in a dual-cell atomic self-oscillating system. Each vapor cell, with a size of $2\times 2\times2$ $\text{cm}^{3}$, is filled with $^{87}\text{Rb}$ atoms and 150 Torr of nitrogen buffer gas. The two cells are positioned inside a five-layer magnetic shield and separated by 10 cm. Independent background magnetic fields are generated by two sets of three-axis coils surrounding each cell, with individually controllable input currents. A circularly polarized pump beam, tuned near the $^{87}\text{Rb}$ $D_1$ transition, polarizes the atomic spins in both ensembles. The spin polarizations of the two cells can be independently adjusted by varying their pump laser intensities. In practice, the pump powers are carefully tuned to achieve nearly identical spin polarizations. A single linearly polarized probe beam, far detuned from the $^{87}\text{Rb}$ $D_2$ transition,  sequentially passes through both vapor cells to measure the total transverse magnetization component $M_x$. Before entering the cells, the probe beam is expanded to a 1 cm diameter using a pair of convex lenses. A variable resistor box connects the photodetector output to the $y$-axis feedback coils, allowing adjustment of the feedback coefficient $\alpha$. This configuration produces a common feedback magnetic field $B_y(t) = -\alpha {M_x}(t) /\gamma$ for both cells.


\subsection{Dynamical behaviors}
\begin{figure*}
    \centering
    \includegraphics[width=1\textwidth]{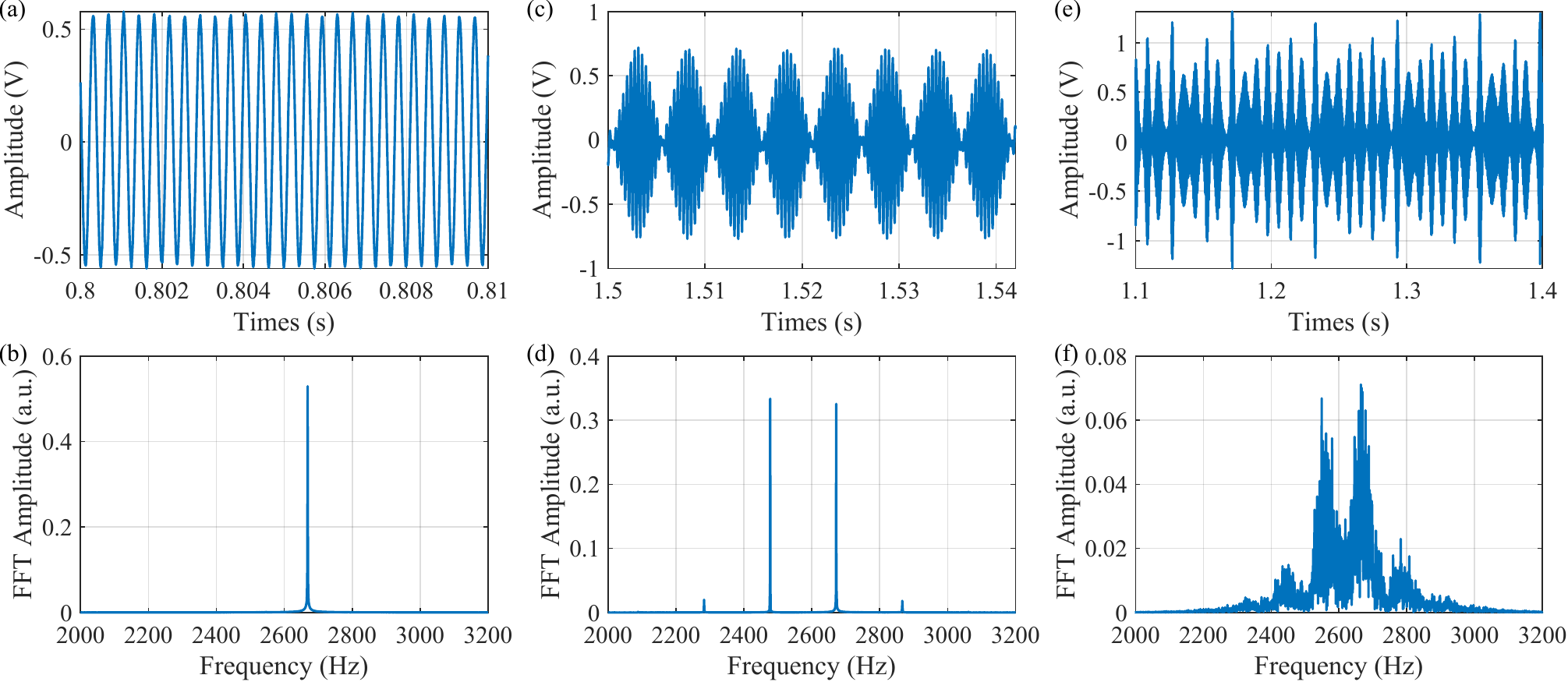} 
    \caption{Experimental results. Time traces and Fourier spectra showing limit-cycle (a,b), quasi-periodic (c,d), and chaotic (e,f) dynamics of the dual-cell self-oscillating system.}
    \label{ExpSpectrumFig}
\end{figure*}
We experimentally observed the time- and frequency-domain signatures of three distinct dynamical behaviors.
The strength of the feedback field is controlled by adjusting the resistance 
$R$ of a variable resistor. A larger 
$R$ corresponds to a smaller feedback coefficient, so 
$1/R$ can be used as a parameter to characterize the feedback coefficient. By varying the frequency difference between the two cells $\Delta\omega$ and the feedback field strength $1/R$, we recorded the probe signal time series under different parameter settings. The corresponding Fourier spectra, shown in Fig.~\ref{ExpSpectrumFig}, were obtained by applying a fast Fourier transform (FFT) to the measured data. With a frequency difference of 40 Hz and a feedback resistance of $6~\mathrm{k}\Omega$, the time trace exhibits periodic oscillations with a stable amplitude, and the Fourier spectrum displays a single sharp peak [Fig.~\ref{ExpSpectrumFig}(a,b)], characteristic of a limit-cycle phase. The linewidth of the peak is limited solely by the finite acquisition time. When the frequency difference is increased to 
230 Hz while maintaining the same resistance, the time-domain signal becomes a sequence of evenly modulated wave packets, and the spectrum shows two dominant peaks at $2.477$ kHz and $2.672$ kHz, accompanied by equally spaced minor sidebands [Fig.~\ref{ExpSpectrumFig}(c,d)]. This spectral structure is the hallmark of a quasi-periodic phase arising from two incommensurate frequencies. The feedback coefficient determines the system’s degree of nonlinearity, with stronger feedback driving the system toward chaos. Reducing the resistance to $3~\mathrm{k}\Omega$ (thereby increasing $\alpha$) and setting $\Delta\omega = 120$ Hz leads to chaotic dynamics. As shown in Fig.~\ref{ExpSpectrumFig}(e,f), the time-domain oscillations exhibit irregular amplitudes, while the Fourier spectrum evolves from discrete peaks into a broad continuous distribution, indicating multiple irregular frequency components and the absence of a well-defined fundamental frequency.


To quantitatively verify the presence of chaos, we applied the chaos decision tree algorithm to the experimentally acquired time-series data. This algorithm outputs a statistical parameter $K$, which characterizes the degree of chaotic behavior in the system. Values of 
$K$ approaching 1 indicate strongly chaotic dynamics, while values near 0 correspond to periodic motion. For the chaotic sequence shown in Fig.~\ref{ExpSpectrumFig}(f), the obtained $K$ value is 0.9923, confirming pronounced chaotic behavior. In contrast, the limit-cycle phase in Fig.~\ref{ExpSpectrumFig}(b) yields $K\approx 0.0623$, and the quasi-periodic phase in Fig.~\ref{ExpSpectrumFig}(d) gives $K\approx 0.2145$, consistent with their expected dynamical characteristics.

Within a broad and continuous parameter range, distinct dynamical behaviors emerge as the system parameters vary. In the following, we fix one parameter while slightly tuning the other to record the system’s phase portraits as functions of the frequency difference $
\Delta f=\Delta \omega /2\pi 
$ and the inverse resistance $1/R$, as shown in Fig.~\ref{PhaseDiagram}. The diagram is composed of numerous coordinate points \((\Delta f, 1/R)\), each corresponding to a time-series signal detected by the photodetector. The white region denotes the no-signal area, where the system lacks sufficient feedback gain to satisfy the self-oscillation condition, and no effective self-excited signal is generated.
The blue region represents the limit-cycle regime, in which the system exhibits stable periodic oscillations. Its spectrum contains a single narrow peak located between the two fixed Larmor frequencies. As seen in the diagram, within a frequency difference range of approximately 70 Hz, the system maintains periodic motion even with increased gain, demonstrating excellent stability. The purple region corresponds to the quasi-periodic regime, characterized by multiple equally spaced spectral lines. Due to the continuous nature of signal acquisition, an intermediate transition zone appears between the no-signal, limit-cycle, and quasi-periodic regions, indicated by the gray area in the diagram.


\begin{figure}
    \centering
\includegraphics[width=0.45\textwidth]{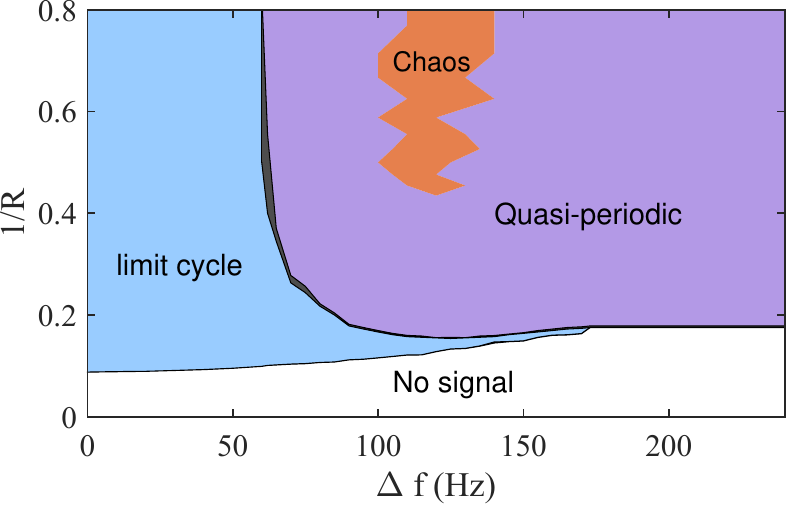} 
    \caption{Experimental phase diagram of the dual-cell system showing regions of no signal, limit cycle (blue), quasi-periodic (purple), and chaotic (orange) dynamics as functions of $\Delta\omega$ and feedback strength $1/R$.}
    \label{PhaseDiagram}
\end{figure}

The distinction between the limit-cycle and quasi-periodic states is determined by the number and positions of spectral lines in the frequency domain. When only a single spectral line is present and its frequency lies between the two intrinsic Larmor frequencies, the system is identified as being in a limit-cycle state; otherwise, it is classified as quasi-periodic or chaotic. Because identifying chaotic behavior requires algorithmic analysis, the boundary between quasi-periodic and chaotic regimes is primarily determined using the 
$K$ value obtained from the chaos decision tree algorithm. Specifically, data sets exhibiting multiple spectral lines are first extracted, and each column of time-series data is analyzed using the 0–1 chaos test. When the resulting 
$K$ value exceeds a defined threshold, the system is identified as chaotic; otherwise, it corresponds to quasi-periodic motion.
The yellow region in the diagram represents points identified as chaotic. It is worth emphasizing that the system demonstrates excellent stability: as the parameters are varied, the system transitions cyclically among the different dynamical regimes in the order indicated by the connecting lines, confirming its strong robustness against external noise perturbations.


\subsection{Robustness analysis}
To further examine the robustness of the system, white noise with an amplitude of 2 V was deliberately introduced into the background magnetic field along the 
$z$-axis. Two representative parameter settings were tested: the limit-cycle regime ($\Delta f = 40$ Hz, $1/R = 0.3$) and the quasi-periodic regime ($\Delta f = 180$ Hz, $1/R = 0.3$). The corresponding experimental results are shown in Fig.~\ref{RobustnessFig}(a,b). As illustrated, the introduction of noise partially disturbs the signals, producing small side peaks in the lower part of the spectrum and reducing the Fourier amplitude of the main peak. Nevertheless, both signals largely preserve the patterns observed in the noise-free case, confirming the system’s inherent stability. In particular, the quasi-periodic motion develops multiple spectral lines as the noise gradually dominates the original main peak, reflecting a typical transition from order to disorder. Remarkably, the limit-cycle regime demonstrates strong robustness: under the same noise perturbation, its spectrum remains dominated by a single sharp peak, closely resembling the noise-free condition.


\begin{figure}
    \centering
    \includegraphics[width=0.45\textwidth]{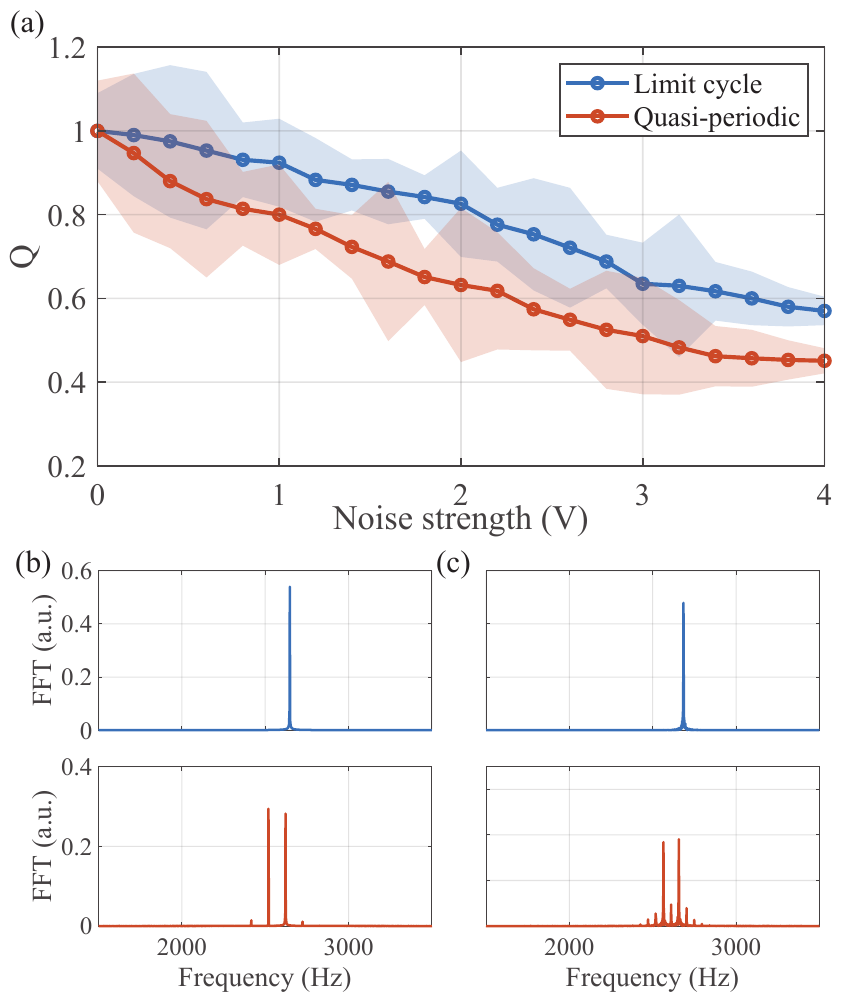} 
    \caption{Noise robustness of limit-cycle and quasi-periodic dynamical regimes. (a) Normalized noise-resistance parameter Q as a function of noise strength. (b) and (c) show the Fourier spectra without and with white noise added. The added noise strength is $2$ V.}
    \label{RobustnessFig}
\end{figure}

To quantitatively evaluate the noise resilience of different dynamical phases, we compared the limit-cycle and quasi-periodic motions under various noise levels and introduced the following metric:
\begin{equation}
Q  = \frac{\int  | A_0(\omega) A_\sigma(\omega)| d\omega/2\pi}{\sqrt{\int | A_0(\omega)|^2 d\omega/2\pi \int |A_\sigma(\omega')|^2 d\omega'/2\pi}}.
	\label{eq: robust}
\end{equation}
where \(A_{\sigma}(\omega)\) denotes the Fourier amplitude of the demodulated signal with added noise strength $\sigma$. At low noise levels, the spectral energy is concentrated near \(\omega_0\), resulting in a large $Q$ value since most of the oscillation energy remains confined around the crystalline frequency. As the noise intensity increases, random fluctuations redistribute the spectral energy over a broader frequency range, reducing the spectral weight near \(\omega_0\) and thus lowering $Q$. This analysis was applied to signals obtained under different noise amplitudes, and the results are shown in Fig.~\ref{RobustnessFig}(c). The horizontal axis represents the amplitude of the applied white noise, while the vertical axis shows the normalized
$Q$ value, quantifying the system’s robustness to noise. Vertical error bars indicate the standard deviation over ten independent measurements. The blue and red curves correspond to the limit-cycle and quasi-periodic phases, respectively.
As the noise intensity increases, $Q$ decreases for both phases. When $Q$ drops to 0.6, the corresponding noise amplitudes are 2.3 V for the quasi-periodic phase and 3.6 V for the limit-cycle phase, demonstrating that the limit-cycle regime exhibits greater robustness against noise perturbations.

}

\section{Conclusions}
In summary, we have explored the nonlinear spin dynamics of a dual-cell self-oscillating rubidium magnetometer, in which two intrinsic Larmor frequencies are coupled through a common feedback loop. This work bridges the gap between theoretical predictions and experimental realizations beyond the single-cell level. Both numerical simulations and experimental observations reveal a rich dynamical phase diagram featuring synchronized limit cycles, quasi-periodic oscillations, and chaotic trajectories. Crucially, the self-sustained oscillations in the limit-cycle and quasi-periodic regimes persist regardless of initial conditions and exhibit exceptional robustness against external perturbations, with the limit-cycle phase showing the highest noise tolerance.
Beyond their fundamental connection to continuous time crystals and time quasi-crystals, these results demonstrate the potential of multifrequency spin masers for high-precision applications such as frequency standards, quantum metrology, and noise-resilient sensing in realistic environments. The dual-cell platform established here provides a versatile testbed for studying complex nonlinear phenomena and advancing practical quantum technologies.


\section*{Acknowledgements}
H.W. was supported by Zhejiang Provincial Natural Science Foundation of China under Grant No. LY24A050005, the Fundamental Research Funds for the Zhejiang University of Science and Technology under Grant No. 2023JLZD010. W.Z. was supported by the National Natural Science Foundation of China under Grant No. 12574532.
Z.L. was supported by the Guangdong Basic and Applied Basic Research Foundation (Grant No. 2024A1515011406), Fundamental Research Funds for the Central Universities, Sun-Yat-Sen University (Grant No. 23qnpy63), and Guangdong Provincial Key Laboratory (Grant No. 2019B121203005)

\section*{References}
\twocolumngrid       


\bibliographystyle{apsrev4-2}
\bibliography{Ref_maser_v2}

\end{document}